\documentclass[conference]{IEEEtran}
\IEEEoverridecommandlockouts
\usepackage{cite}
\usepackage{amsmath,amssymb,amsfonts}
\usepackage{algorithmic}
\usepackage{graphicx}
\usepackage{textcomp}
\makeatletter

\newcommand{\Rmnum}[1]{\expandafter\@slowromancap\romannumeral #1@}
\makeatother
\usepackage{multirow, makecell}
\def\BigRoman{\uppercase\expandafter{\romannumeral\number\count 255 }}
\def\Romannumeral{\afterassignment\BigRoman\count255=}
\usepackage{tikz}
\def\BibTeX{{\rm B\kern-.05em{\sc i\kern-.025em b}\kern-.08em
    T\kern-.1667em\lower.7ex\hbox{E}\kern-.125emX}}

\begin{document}

\title{Decoding Multi-class Motor-related Intentions with User-optimized and Robust BCI System Based on Multimodal Dataset 
\footnote{\thanks{This work was partly supported by Institute of Information \& Communications Technology Planning \& Evaluation (IITP) grant funded by the Korea government (MSIT) (No. 2017-0-00432, Development of Non-Invasive Integrated BCI SW Platform to Control Home Appliances and External Devices by User’s Thought via AR/VR Interface; No. 2017-0-00451, Development of BCI based Brain and Cognitive Computing Technology for Recognizing User’s Intentions using Deep Learning; No. 2019-0-00079, Artificial Intelligence Graduate School Program, Korea University).}
}
}

\author{\IEEEauthorblockN{Jeong-Hyun Cho}
\IEEEauthorblockA{\textit{Dept. Brain and Cognitive Engineering}\\
\textit{Korea University} \\
Seoul, Republic of Korea \\
jh\_cho@korea.ac.kr}\\

\and

\IEEEauthorblockN{Byoung-Hee Kwon}
\IEEEauthorblockA{\textit{Dept. Brain and Cognitive Engineering}\\
\textit{Korea University} \\
Seoul, Republic of Korea \\
bh\_kwon@korea.ac.kr}\\

\and

\IEEEauthorblockN{Byeong-Hoo Lee}
\IEEEauthorblockA{\textit{Dept. Brain and Cognitive Engineering}\\
\textit{Korea University} \\
Seoul, Republic of Korea \\
bh\_lee@korea.ac.kr}\\
}


\maketitle

\begin{abstract}
A brain-computer interface (BCI) based on electroencephalography (EEG) can be useful for rehabilitation and the control of external devices. Five grasping tasks were decoded for motor execution (ME) and motor imagery (MI). During this experiment, eight healthy subjects were asked to imagine and grasp five objects. Analysis of EEG signals was performed after detecting muscle signals on electromyograms (EMG) with a time interval selection technique on data taken from these ME and MI experiments. By refining only data corresponding to the exact time when the users performed the motor intention, the proposed method can train the decoding model using only the EEG data generated by various motor intentions with strong correlation with a specific class. There was an accuracy of 70.73\% for ME and 47.95\% for MI for the five offline tasks. This method may be applied to future applications, such as controlling robot hands with BCIs.
\end{abstract}

\begin{small}
\end{small}

\section{Introduction}
A brain-computer interface (BCI) is a method of analyzing brain signals in order to determine a person's intent and status, which can then be used for analyzing various motor imagery (MI). The purpose of a lot of BCI studies is to attempt to understand brain signals, as brain signals can contain significant information about a person's status \cite{C3,MRCP,B1,ECoG2}. As an alternative, invasive methods such as electrocroticogram (ECoG) \cite{ECoG} are used to obtain high-quality brain signals by placing electrodes directly on the brain. In comparison with non-invasive methods like electroencephalograms (EEG), functional near-infrared spectroscopy, and functional magnetic resonance imaging that can be used to obtain brain signals, these techniques are more risky because the electrodes have to be implanted during surgery. For the purposes of further advancing BCI research using EEG, researchers are using different EEG paradigms to detect user intentions and applying the results to different applications. For instance, EEG-based BCI has several paradigms for signal acquisition such as MI \cite{A2,C2,kam}, event-related potential (ERP) \cite{EEG,A1}, and movement-related cortical potential \cite{jeong2020decoding,MRCP}. As applications of EEG-based BCI, a systemic arm \cite{A2}, a speller \cite{speller,won2017motion}, a wheelchair \cite{wheelchair}, and neuroprosthetics \cite{osborn2018prosthesis, ganzer2020restoring, tayeb2020decoding} were commonly used for communication between human and machines. Our research focused on ME and MI decoding for applications in neuroprosthetics, such as robotic arm control, and devised a user-optimized data editing technique to enhance model learning efficiency. 

\begin{figure*}[t!]
\centerline{\includegraphics[width=\textwidth]{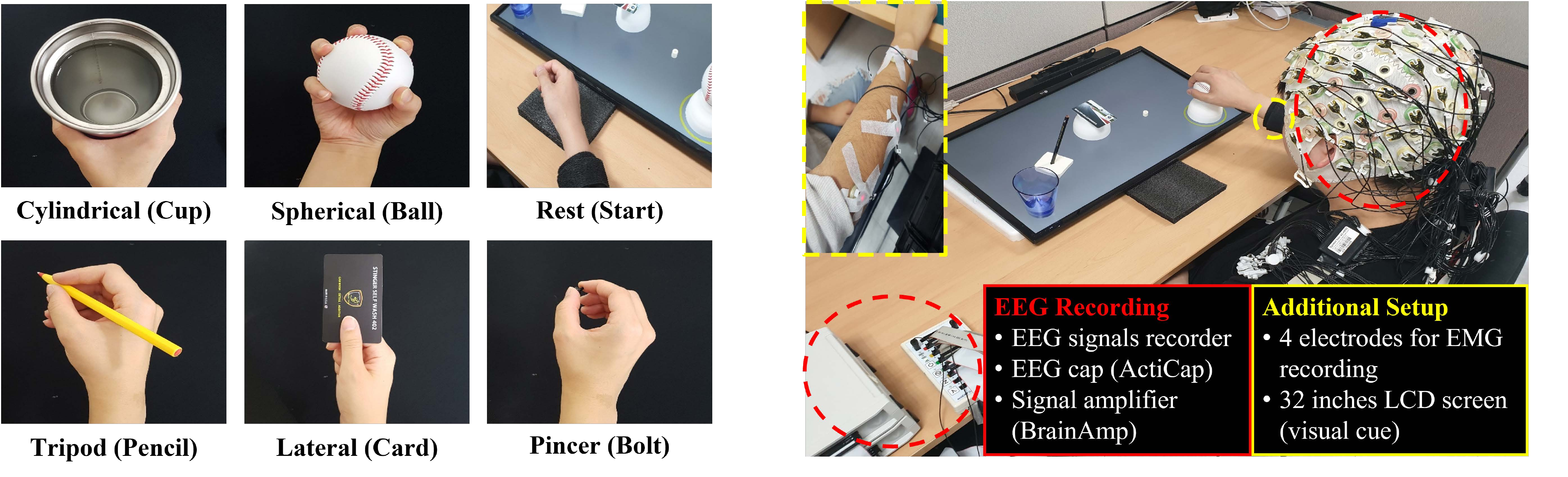}}
\caption{(left) Five different reach-and-grasp tasks. Each grasping action is associated with a specific object. (right) Experimental environment for acquiring EEG and EMG signals.}
\end{figure*}

Our research direction was determined by several similar previous studies. Schwarz \emph{et al}. \cite{schwarz2017decoding} decoded natural reach and grasp actions using human EEG data. The authors used EEG neural correlates to identify three different grasps and reach actions: lateral grasp, pincer grasp, and palmar grasp. Their offline ME experiment yielded an accuracy of 74.2\% for binary classification of grasp types and 65.9\% for multiclass classification of grasp types. In the other case, Ofner \emph{et al}. \cite{ofner2017upper} encoded one upper limb movement in the time domain of low-frequency EEG signals. Five different movements were classified in this experiment: elbow flexion, extension, grasp, spread, wrist twist left, and wrist twist right. Using a different approach, Agashe \emph{et al}. \cite{agashe2015global} decoded hand motions with a final classification accuracy of 55\% in ME and 27\% in MI. They discovered that global cortical activity predicts grasping hand shapes. As subjects made natural reach-to-grasp movements, EEG and hand joint angular velocity were determined as well as synergistic trajectory estimates. A closed-loop neuroprosthetic system was used to help an amputee grasp in real time, and brain signals could be decoded to decode various hand motions. 

We present a non-invasive BCI methodology for assessing grasping tasks based on EEG signals in this paper. In addition, we found that EEG signals could be used to classify grasping tasks performed by the right hand by analyzing them. By conducting enough experiments and analyzing the data, we built a robust decoding model that was built on the data. As a result of this model, robotic hand control systems based on MI and driven by EEG signals may be developed in the future.

\begin{figure}[t!]
\centerline{\includegraphics[width=\columnwidth]{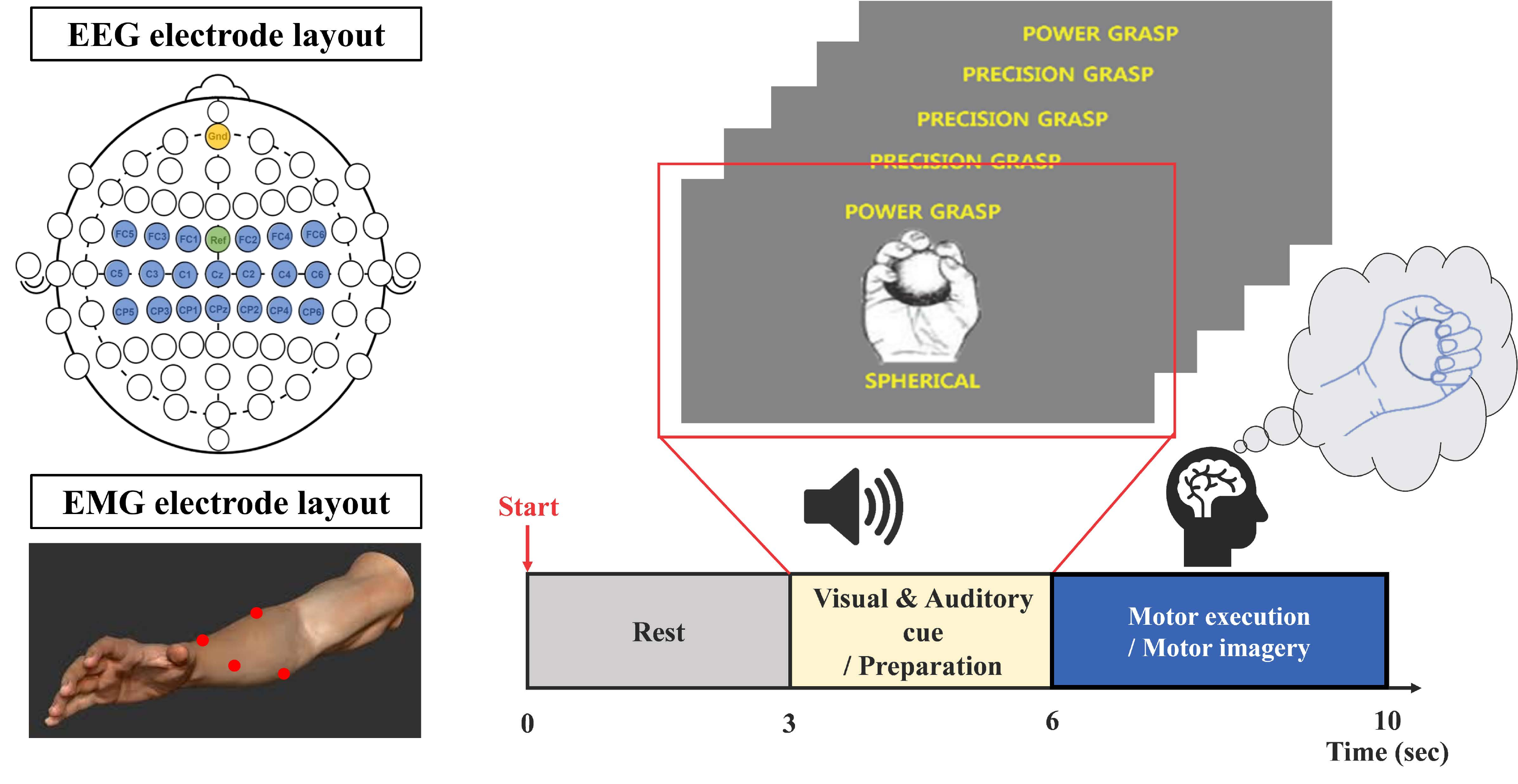}}
\caption{Illustration of (left) electrodes locations and (right) experimental protocol to decode five class reach-and-grasp actions in actual movements and motor imagery paradigm.}
\end{figure}

\section {Materials and Methods}
\subsection{Participants}
Our study recruited eight healthy individuals without any history of neurological disease (age 24--29, right-handed males), and we purposely selected all subjects who had already participated in the BCI experiment. As part of our effort to achieve the finest possible results and to make sure that the decoding of complex MI tasks was feasible, we collected data from trained individuals \cite{C2, C4, jeong2020brain, zhang2017hybrid}. The study was reviewed and approved by the International Review Board at Korea University [1040548-KU-IRB-17-172-A-2] and written informed consent was obtained from each participant before the study could proceed.

\begin{figure*}[t]
\centerline{\includegraphics[width=\textwidth]{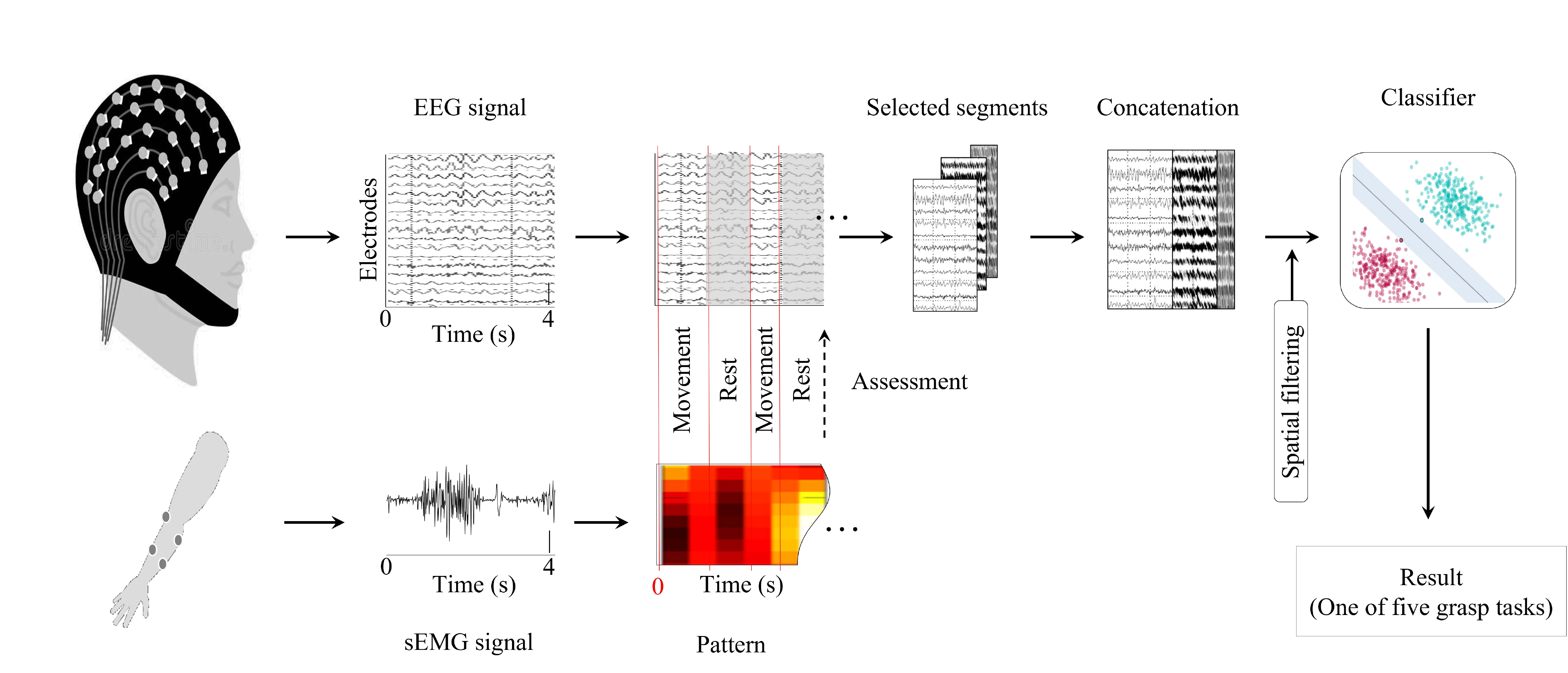}}
\caption{An approach to selecting subject-specific time segments during feature selection based on EMG signals. Measuring EMG signals to determine the optimal time segments for feature selection. From 0 to 4 seconds, choose a specific length segment (e.g., 1 second or 2.5 seconds).}
\end{figure*}

\subsection{Experimental Setup}
Subjects were seated comfortably in a comfortable chair during each session of the experimental protocol in front of a 32-inch horizontally installed LCD monitor screen, as depicted in Figure 1. Upon completion of each auditory and visual cue, subjects were asked to visualize or perform a specific grasping task. During the rest period, the subjects mimicked the designated grasping task, including a half-opened hand shape which illustrates hand relaxation during the grasping process. As shown in figure 1, the subjects were asked to perform five different grasping tasks.

During the experiment, the EEG data were collected at 1000 Hz using 64 Ag/AgCl electrodes in 10/20 international system via BrainAmp (BrainProduct GmbH). The FCz and FPz were used as reference and ground electrodes, respectively. From the entire 64 EEG channels, only 20 channels (FC5, FC3, FC1, FC2, FC4, FC6, C5, C3, C1, Cz, C2, C4, C6, CP5, CP3, CP1, CPz, CP2, CP4, and CP5) were used for data processing \cite{kwon2019subject, kim2019subject}, as shown in figure 2. The 20 channels were located only on the motor cortex to make sure that the recorded EEG signals are highly related to the motor-related potentials, which are from the ME and MI \cite{MRCP, MI, nonstationary, won2017motion, kam}. 

This experiment consisted of 50 trials for each grasping task. Furthermore, each of the subjects was asked to handle their hands in a particular way once but not repeatedly during the ME/MI period of four seconds.     

\subsection{EEG Signal Acquisition}
A BrainVision Recorder was used to acquire the EEG signals (BrainProducts GmbH, Germany). The EEG signals were collected using 64 Ag/AgCl electrodes which followed 10/20 international protocols. The reference and ground channels were respectively FCz and FPz positions. We sampled the signals at 1,000 Hz and applied a 60 Hz notch filter to the signals that were acquired. All electrode impedances were kept below 10 k$\Omega$ during the experiment. 

In addition to these motor-related bands, movement-related cortical potentials (MRCPs) [0.3-3] Hz were also taken into account in order to determine the relationship between classifying MI and readiness potentials \cite{FBCSP, SSVEP2, MRCP, book}. We have already identified this idea in a previously conducted study \cite{averagepooling}. For this experiment, we used the [0.3-30] Hz frequency band because it showed the most reliable classification result compared to the other bands. Muscle-related signal changes were detected through analysis of the EMG signal. In our analyses, we assumed that when the root-mean-square (RMS) EMG was over the threshold level, the subject was doing a grasping task.

\subsection{Data Analysis}
EEGNet, which has demonstrated outstanding performance in classification tasks, along with linear discriminant analysis (LDA) were compared against the algorithm for common spatial patterns (CSP). Our data segmentation and concatenation process allowed us to prepare a 4-second dataset for training with the CSP algorithm as a result of dividing each trial into various sizes. \cite{FBCSP, A3}. In order to construct the feature, a transformation matrix consisting of the logarithmic variances of the first and last columns of the CSP was created. Using the LDA \cite{channel, CNN, MI2}, five different classes were classified according to the one-versus-rest classification \cite{intuitiveMI, highertemporal, wheelchair, ADAM}. 

The preprocessed EEG data were used in the training process for EEGNet, an architecture based on a convolutional neural network (CNN). Despite the fact that conventional feature extraction methods such as CSP were not implemented, only significant features were likely to be extracted with the use of the convolution layer. A ten-by-ten cross-validation test was carried out with the objective of evaluating classification performance \cite{cho2020decoding, cho2020novel, channel}.

As a final step, we analyzed the decoding performance of EEGNet trained with training data that did not undergo any data editing other than preprocessing. In contrast, another EEGNet trained with training data that had been edited using the proposed method that can generate a data set optimized for its users. 

\renewcommand{\arraystretch}{1.25}
\begin{table}[t!]
\centering
\caption{The accuracy of 5-class classification to be achieved in ME and MI paradigms using data editing technologies that are optimized for the use of multimodal data}
\label{table1}
\resizebox{\columnwidth}{!}{%
\begin{tabular}{ccccc}
\hline
\multicolumn{1}{l}{\multirow{3}{*}{\textbf{}}} &
  \multicolumn{4}{c}{\textbf{\begin{tabular}[c]{@{}c@{}}5-Class classification accuracy\\ (10-by-10-fold cross-validation)\end{tabular}}} \\ \cline{2-5} 
\multicolumn{1}{l}{} &
  \multicolumn{2}{c}{\textbf{ME}} &
  \multicolumn{2}{c}{\textbf{MI}} \\
\multicolumn{1}{l}{} &
  \textbf{Conventional} &
  \textbf{Proposed} &
  \textbf{Conventional} &
  \textbf{Proposed} \\ \hline
\textbf{Sub 1} &
  \begin{tabular}[c]{@{}c@{}}0.6770\\ ($\pm$0.0294)\end{tabular} &
  \begin{tabular}[c]{@{}c@{}}0.8338\\ ($\pm$0.1021)\end{tabular} &
  \begin{tabular}[c]{@{}c@{}}0.5125\\ ($\pm$0.0350)\end{tabular} &
  \begin{tabular}[c]{@{}c@{}}0.5552\\ ($\pm$0.0433)\end{tabular} \\
\textbf{Sub 2} &
  \begin{tabular}[c]{@{}c@{}}0.5140\\ ($\pm$0.0503)\end{tabular} &
  \begin{tabular}[c]{@{}c@{}}0.7787\\ ($\pm$0.0935)\end{tabular} &
  \begin{tabular}[c]{@{}c@{}}0.4601\\ ($\pm$0.1305)\end{tabular} &
  \begin{tabular}[c]{@{}c@{}}0.4117\\ ($\pm$0.0637)\end{tabular} \\
\textbf{Sub 3} &
  \begin{tabular}[c]{@{}c@{}}0.5380\\ ($\pm$0.1113)\end{tabular} &
  \begin{tabular}[c]{@{}c@{}}0.6227\\ ($\pm$0.1208)\end{tabular} &
  \begin{tabular}[c]{@{}c@{}}0.3475\\ ($\pm$0.0751)\end{tabular} &
  \begin{tabular}[c]{@{}c@{}}0.3933\\ ($\pm$0.0116)\end{tabular} \\
\textbf{Sub 4} &
  \begin{tabular}[c]{@{}c@{}}0.4875\\ ($\pm$0.0891)\end{tabular} &
  \begin{tabular}[c]{@{}c@{}}0.6873\\ ($\pm$0.0813)\end{tabular} &
  \begin{tabular}[c]{@{}c@{}}0.3525\\ ($\pm$0.0385)\end{tabular} &
  \begin{tabular}[c]{@{}c@{}}0.4930\\ ($\pm$0.0181)\end{tabular} \\
\textbf{Sub 5} &
  \begin{tabular}[c]{@{}c@{}}0.5558\\ ($\pm$0.0699)\end{tabular} &
  \begin{tabular}[c]{@{}c@{}}0.7250\\ ($\pm$0.0732)\end{tabular} &
  \begin{tabular}[c]{@{}c@{}}0.4709\\ ($\pm$0.0407)\end{tabular} &
  \begin{tabular}[c]{@{}c@{}}0.4655\\ ($\pm$0.0974)\end{tabular} \\
\textbf{Sub 6} &
  \begin{tabular}[c]{@{}c@{}}0.6740\\ ($\pm$0.0871)\end{tabular} &
  \begin{tabular}[c]{@{}c@{}}0.6912\\ ($\pm$0.0942)\end{tabular} &
  \begin{tabular}[c]{@{}c@{}}0.3254\\ ($\pm$0.0547)\end{tabular} &
  \begin{tabular}[c]{@{}c@{}}0.3845\\ ($\pm$0.0612)\end{tabular} \\
\textbf{Sub 7} &
  \begin{tabular}[c]{@{}c@{}}0.6832\\ ($\pm$0.1064)\end{tabular} &
  \begin{tabular}[c]{@{}c@{}}0.7312\\ ($\pm$0.0371)\end{tabular} &
  \begin{tabular}[c]{@{}c@{}}0.4003\\ ($\pm$0.1009)\end{tabular} &
  \begin{tabular}[c]{@{}c@{}}0.5439\\ ($\pm$0.1179)\end{tabular} \\
\textbf{Sub 8} &
  \begin{tabular}[c]{@{}c@{}}0.4593\\ ($\pm$0.1064)\end{tabular} &
  \begin{tabular}[c]{@{}c@{}}0.5885\\ ($\pm$0.0371)\end{tabular} &
  \begin{tabular}[c]{@{}c@{}}0.4494\\ ($\pm$0.1009)\end{tabular} &
  \begin{tabular}[c]{@{}c@{}}0.5890\\ ($\pm$0.1179)\end{tabular} \\
\textbf{\begin{tabular}[c]{@{}c@{}}Average\\ ($\pm$Std.)\end{tabular}} &
  \textbf{\begin{tabular}[c]{@{}c@{}}0.5736\\ ($\pm$0.0913)\end{tabular}} &
  \textbf{\begin{tabular}[c]{@{}c@{}}0.7073\\ ($\pm$0.0792)\end{tabular}} &
  \textbf{\begin{tabular}[c]{@{}c@{}}0.4148\\ ($\pm$0.0682)\end{tabular}} &
  \textbf{\begin{tabular}[c]{@{}c@{}}0.4795\\ ($\pm$0.0786)\end{tabular}} \\ \hline
\end{tabular}
}
\end{table}


\section {Results and Discussion}
Table I presents the classification accuracy results in ME and MI based on the study which demonstrated the feasibility of classifying five grasp tasks in the right hand. In the ME session, every subject showed an increase in accuracy of approximately three to fifteen percent as compared to the conventional method. It can be seen in Table I that classification accuracy for subjects S2 and S5 was significantly improved by segmenting them by 2.5 seconds. Compared to the conventional method, the accuracy of the other subjects was slightly higher with a peak of 83.38\%. The classification accuracy for some of the subjects decreased in the MI session, whereas in the MI session, classification accuracy increased continuously. As a result of applying our proposed decoding method, the majority of the subjects achieved higher classification accuracy. In the case of S2 and S5, classification accuracy started to decrease during the MI paradigm. However, the subjects showed a small decrease while the other subjects showed a more significant increase in accuracy. Similarly, the S4 and S7 achieved dramatic improvements of 13\% and 14\%, respectively. 

A different time point in each trial was assigned to each subject. At the same time, they performed the ME for segmentation of their EEG data. In the ME session, subjects were required to move their body parts according to instructions in real time. This resulted in lower levels of performance variation than during the MI session. There is a strong correlation between subject-specific time interval selections and the different results for five-class multi-classification accuracy between actual movement and MI. In comparison to using the same method with the actual movement paradigm, the proposed method of detecting muscle-related signals by referring EMG signals to the MI paradigm demonstrated a more significant increase in classification accuracy. As an example, the proposed method was able to improve accuracy from 5 to 14\% during MI classification. This is compared to the comparison model which used an ordinary fixed 0 to 4,000 ms time interval.

Our study confirms the effectiveness of our proposed subject-specific time interval selection method in improving classification accuracy both in a MI paradigm as well as in a movement paradigm using ME data. Based on the results, the proposed method can be utilized as a feature extraction strategy to improve multi-class classification accuracy on a set purpose, especially in the MI paradigm. Despite the fact that the experiment was conducted in an offline scenario using a visual cue-based synchronous system, subjects completed MI at various speeds, trajectories, and forces \cite{speller, C2}. A subject only used intuitive MI-BCI when performing specific hand motions during each of the trials. The selection of the time interval length to extract movement-related EEG features during the MI paradigm is useful when the subjects perform variable motor-related brain activity \cite{B2, B1}.  

\section{Conclusion and Future works}
We have confirmed the feasibility of classifying five different grasping tasks based on EEG signals within the hand. Our proposed method improved overall classification accuracy. Our future research will focus on improving the reliability and efficiency of MI decoding algorithms utilizing advanced machine learning and deep learning approaches to develop robust decoding methods. For the development of practical BCI-based robot hand control, we will also modify our offline experimental protocol to be asynchronous and online oriented \cite{roboticarm, cho2020decoding, jeong2020decoding}.

As a consequence, the EEG classification performance needs to be higher. As part of our ongoing efforts to make our motor-related BCI system more robust in real-world settings, we will continue to test and implement advanced deep learning approaches. This would significantly improve the interaction effect between the user and the BCI.

\section*{Acknowledgment}
The authors thank to M.-K. Kim and H.-B. Shin for help with the database construction and useful discussions of the experiment. 

\bibliographystyle{IEEEbib}
\bibliography{refs}

\end{document}